\documentclass[aps,pra,showpacs,superscriptaddress,twocolumn,final,numerical,10pt]{revtex4-1}
\usepackage{epsfig}
\usepackage{graphicx}
\usepackage{dcolumn}
\usepackage{bm}
\usepackage{amsthm}
\usepackage{amsfonts}
\usepackage{amscd}
\usepackage{amsmath}  
\usepackage{amssymb}
\usepackage{enumerate}
\usepackage{color}

\newcommand{\ket}[1]{\mbox{$ | #1 \rangle $}}
\newcommand{\bra}[1]{\mbox{$ \langle #1 | $}}

\begin{document}

\title{Security of quantum key distribution with iterative sifting}

\author{Kiyoshi Tamaki}
\email{tamaki@eng.u-toyama.ac.jp}
\affiliation{Department of Intellectual Information Engineering, Faculty of Engineering,
University of Toyama, Gofuku 3190, Toyama 930-8555, Japan}
\affiliation{NTT Basic Research Laboratories, NTT Corporation, 3-1,Morinosato Wakamiya Atsugi-Shi, Kanagawa, 243-0198, Japan}
\author{Hoi-Kwong Lo}
\affiliation{Center for Quantum Information and Quantum Control, Dept. of Electrical \& Computer Engineering and Dept. of Physics,
University of Toronto, M5S 3G4, Canada}
\author{Akihiro Mizutani}
\affiliation{Graduate School of Engineering Science, Osaka University, Toyonaka, Osaka 560-8531, Japan}
\author{Go~Kato}
\affiliation{NTT Communication Science Laboratories, NTT Corporation, 3-1,Morinosato Wakamiya Atsugi-Shi, Kanagawa, 243-0198, Japan}
\author{Charles Ci Wen Lim}
\affiliation{Group of Applied Physics, University of Geneva, Geneva CH-1211, Switzerland}
\thanks{C. C. W. Lim's current affiliation is ``Quantum Information Science Group, Computational Sciences and Engineering Division,
Oak Ridge National Laboratory, Oak Ridge, Tennessee 37831-6418, US''.}
\author{Koji Azuma}
\affiliation{NTT Basic Research Laboratories, NTT Corporation, 3-1,Morinosato Wakamiya Atsugi-Shi, Kanagawa, 243-0198, Japan}
\author{Marcos Curty}
\affiliation{EI Telecomunicaci\'on, Dept. of Signal Theory and Communications, University of Vigo, E-36310, Spain}

\begin{abstract}
Several quantum key distribution (QKD) protocols employ iterative sifting.
After each quantum transmission round, Alice and Bob disclose part of their setting
information (including their basis choices) for the detected signals. This quantum
phase then ends when the basis dependent termination conditions are met, i.e., the numbers of detected signals per basis exceed
certain pre-agreed threshold values. Recently, however, Pfister {\it et al}. [New J. Phys.
{\bf 18} 053001 (2016)] showed that the basis dependent termination condition makes QKD insecure, especially in the finite key regime, and
they suggested to disclose all the setting information after finishing the quantum phase. However, this protocol has two main drawbacks: it requires that Alice
possesses a large memory, and she also needs to have some a priori
knowledge about the transmission rate of the quantum channel.

Here we solve these two problems by introducing a basis-independent
termination condition to the iterative sifting in the finite key regime.
The use of this condition, in combination with Azuma's inequality,
provides a precise estimation on the amount of privacy amplification that
needs to be applied, thus leading to the security of QKD protocols,
including the loss-tolerant protocol [Phys. Rev. A 90, 052314 (2014)],
with iterative sifting. Our result allows the implementation of wider
classes of classical post-processing techniques in QKD with quantified
security.

\end{abstract}
\maketitle

\section{Introduction}

Quantum key distribution (QKD) is a way to distribute a key to distant parties in an information-theoretic secure manner, 
and this key is mainly to be used in one-time pad to achieve secure communication. Conventionally, QKD protocols,
such as the BB84 scheme
\cite{BB84}, run the following two phases alternately and repeatedly: a quantum communication phase and a classical post-processing phase.
In the former phase, the sender (Alice) randomly chooses a basis to encode a bit value into an optical
pulse that she sends to the receiver (Bob). On the receiving side, Bob randomly chooses a
measurement basis to decode the bit value encoded in the optical pulse received.
After each quantum communication round, Alice and Bob move on to the latter phase where
they usually employ an
authenticated public channel to disclose part of their information, which includes
their basis choices and the information on whether or not Bob detected the pulse. The quantum communication phase
terminates when the basis dependent termination conditions are satisfied, i.e., when the
numbers of the basis-matched detection instances for each basis exceed certain pre-agreed threshold values.
This procedure is called iterative sifting, and is widely used in several QKD
protocols and security proofs \cite{BB84 random}.

Recently, Pfister {\it et al}. \cite{iterative} showed that, rather surprisingly, when the BB84 protocol uses the basis dependent iterative sifting,
especially in the finite key regime, it is insecure if one 
employs the random sampling theory for the parameter estimation for privacy amplification.
In so doing, they noticed that the basis dependent iterative sifting violates the random sampling
assumption because Eve's knowledge of Alice and Bob's iterative sifting discussion allows her to
bias the sampling. Moreover, it violates basis-independence because the termination condition
is basis-dependent.
This insecurity is a serious problem because it implies that a number of existing finite key security proofs \cite{BB84 random} could be flawed and,
therefore, cannot guarantee the security of the BB84 protocol with iterative sifting.
To solve this issue, Pfister {\it et al}. \cite{iterative} suggested to postpone all the announcement until the end of the quantum communication phase,
as Lo, Chau and Ardehali \cite{random sampling, LCA} proposed. That is, the idea is to fix the number of pulses sent rather than the number of pulses detected. 

The insecurity of the iterative sifting means in particular that Alice needs a large memory
to store all the basis and bit information until the end of the quantum communication phase, which we would like to avoid in practice.
Moreover, in order to properly determine the number of pulses to be sent, Alice has to have a good a priori knowledge on the transmission rate
of the quantum channel, which is difficult because Eve may change her strategy frequently.
Hence, it is very important for the field of QKD to find a solution to these problems. Also, such a solution
can shed new light on how to deal with information disclosure in QKD schemes
without threatening their security.

In this paper, we address this issue and present a simple proof to show that if 
a QKD protocol employs the basis independent termination condition for iterative sifting, instead of the basis dependent one,
and if one employs
Azuma's inequality \cite{Azuma}, which is more general than the random sampling theory, to estimate the parameters
needed for privacy amplification, then the QKD protocol is secure even in the finite key regime.
Here, note that the quantum communication phase terminates in the basis independent manner, but Alice and Bob
are allowed to announce basis information after each round of the quantum transmission part. 
With this sifting, the memory requirements of Alice and Bob is reduced
because they can simply throw away the bit value when the signals are 
not detected, and they do not need to have a good a priori knowledge on the transmission rate
of the quantum channel.

In order to provide an intuition of  our idea, suppose that
Alice and Bob are at the $i^{th}$ quantum communication round of a QKD scheme with
iterative sifting. That is, Eve has at her disposal all the classical information (which
includes the basis information) announced by Alice and Bob in the previous
$(i-1)$ rounds. This means, in particular, 
that she can adapt her eavesdropping strategy for the $i^{th}$ pulse sent by Alice based on such available 
information \cite{footnote2}. This implies that the random sampling theory cannot be applied to this scenario because,
roughly speaking, this theory requires that Eve's attack on the whole signals sent by Alice
must be independent of all the basis information. Importantly, however, thanks to the basis independence of the
termination condition as well as due to
the fact that Eve does not have any a priori knowledge on the basis choices 
that Alice and Bob will make at each of the rounds,
Eve's strategy for the $i^{th}$ pulse cannot depend on the classical
information that Alice and Bob will announce at the end of that round. 
And it turns out that these two conditions are enough to be able to apply Azuma's inequality \cite{footnote}. As a
consequence, one can prove the security of QKD with iterative sifting. 
We remark the following implication of our result: if the security proof of a QKD protocol employs Azuma's inequality 
for the parameter estimation needed for privacy amplification \cite{Azuma2, Mizutani}, this protocol is still secure even if it adopts
basis independent iterative sifting.
This is so because one of the essential points of such proofs is the independence of the basis choice at each
round, and therefore they can safely adopt basis-independent iterative sifting without compromising
their security. Such protocols include, for example, the loss-tolerant
protocol \cite{LT} and its generalized version \cite{in preparation}.

The paper is organized as follows. In Sec.~\ref{Assumption}, we summarize the assumptions that we make in this paper.
Then, in Sec.~\ref{Actual protocol} we introduce the BB84 protocol with basis independent iterative sifting. 
In Sec.~\ref{Virtual protocol}, we present a virtual protocol that is mathematically equivalent to
the BB84 scheme with basis independent iterative sifting. This virtual protocol will be used in the
security proof that we provide later on. Also, in this section, we explain why the
BB84 protocol with basis dependent iterative sifting is insecure when one employs the random
sampling theory. Then, in Sec.~\ref{Security proof}, we present our main result to show that the protocol
is secure if one uses Azuma's inequality and basis independent iterative sifting. Finally, we discuss the security of the loss-tolerant
protocol with iterative sifting and the basis independent termination condition, and summarize our
findings in Sec.~\ref{Discussion}.

\section{Assumptions and definitions}\label{Assumption}

In this section, we summarize the assumptions and definitions that we use in this paper. For simplicity, we consider the BB84 protocol
with a single-photon source.  
Alice randomly chooses a bit value and encodes it using the {\rm Z} or the {\rm X} basis, which
she selects with probability $p_{\rm Z}^{\rm (A)}$ and $p_{\rm X}^{\rm (A)}$, respectively. We express the single-photon state generated by Alice when she chooses a bit value $b$ and the basis $\alpha$ as
$\ket{b_{\alpha}}_{\rm B}$. 
Here, the subscript ${\rm B}$ means that this single-photon system will be sent to Bob, and we define the ${\rm Z}$ (${\rm X}$) basis as
$\{\ket{0_{\rm Z}}, \ket{1_{\rm Z}}\}$ ($\{\ket{0_{\rm X}}, \ket{1_{\rm X}}\}$) with $\ket{0_{\rm X}}:=(\ket{0_{\rm Z}}+\ket{1_{\rm Z}})/\sqrt{2}$ and
$\ket{1_{\rm X}}:=(\ket{0_{\rm Z}}-\ket{1_{\rm Z}})/\sqrt{2}$. Also, for simplicity, we shall assume that the state preparation is perfect. 

As for Bob,
he measures the incoming system ${\rm B'}$, which is the system ${\rm B}$ after Eve's intervention, using
the ${\rm Z}$ basis or the ${\rm X}$ basis, which he chooses with probability $p_{\rm Z}^{\rm (B)}$ and $p_{\rm X}^{\rm (B)}$, respectively.
Note that the state of the system ${\rm B'}$ is not necessarily
a qubit state. But, we assume that Bob's measurement satisfies the basis independent 
detection efficiency condition. That is, when the ${\rm Z}$ (${\rm X}$) basis measurement is represented by
a positive operator valued measure (POVM) $M_{\rm Z}=\{M_{0{\rm Z}}, M_{1{\rm Z}}, M_{\rm fail}\}$ ($M_{\rm X}=\{M_{0{\rm X}}, M_{1{\rm X}}, M_{\rm fail}\}$), then the operator $M_{\rm fail}$ is the same for both bases \cite{LT}. Here, $M_{b\alpha}$ represents the POVM element
associated to the detection of the bit value $b$ in the basis $\alpha$, and the operator $M_{\rm fail}$
represents the failure of outputting a bit value. We call an instance with an outcome
associated to the operator $M_{b\alpha}$ a detected instance.

In addition, for simplicity, we shall assume that in the error correction step of the
protocol, Alice and Bob agree in advance on a family of error correcting codes to
correct the errors. Afterward, they check if the errors have been corrected by
employing an error verification step, which succeeds with high probability. Finally,
we consider that there are no side-channels available to Eve from Alice and Bob's
labs.

\section{DESCRIPTION OF THE BB84 PROTOCOL WITH BASIS INDEPENDENT ITERATIVE SIFTING}\label{Actual protocol}

In this section, we provide the description of the actual BB84 protocol with basis independent iterative
sifting; it runs as follows.
\newline

(Actual protocol)
\begin{enumerate}
\renewcommand{\labelenumi}{(\arabic{enumi})}
\item Alice and Bob agree, over an authenticated public channel, on a number $N_{\rm det}^{\rm (Ter)}$, the secrecy 
parameter $\epsilon_{\rm s}$, the correctness parameter $\epsilon_{\rm c}$, a family of
error correcting codes, and two families of hash functions, one for privacy
amplification and the other one for error verification. Also, they initialize
a parameter $N_{\rm det}$ to zero, and they announce the probabilities $p_{\rm Z}^{\rm (A)}$ ($p_{\rm Z}^{\rm (B)}$)
and $p_{\rm X}^{\rm (A)}$ ($p_{\rm X}^{\rm (B)}$) with which Alice (Bob) chooses the ${\rm Z}$ and ${\rm X}$ bases, respectively. 
\item (Loop phase 1) Alice chooses the basis $\alpha\in\{{\rm Z}, {\rm X}\}$ with probability $p_{\alpha}^{\rm (A)}$, and she
chooses a bit value $b$ at random. She records these values, and sends
Bob a single-photon state $\ket{b_{\alpha}}_{\rm B}$.
\item (Loop phase 2) Bob measures the incoming system ${\rm B'}$ using the {\rm Z} ({\rm X}) basis,
which he selects with probability $p_{\rm Z}^{\rm (B)}$ ($p_{\rm X}^{\rm (B)}$). Then he informs Alice over an authenticated public
channel whether or not he detected the signal as well as the basis information. If the signal is detected, Alice
informs Bob of the basis choice over an authenticated public channel, Alice and Bob increase $N_{\rm det}$ by 1 unit, and 
Bob records his basis choice and the resulting bit value. If $N_{\rm det}=N_{\rm det}^{\rm (Ter)}$ is satisfied, Alice and Bob
proceed to Step (4), otherwise Alice and Bob return to Step (2).
\item (Defining the sifted key) 
Each of Alice and Bob defines the bit string originating from the detected instances with the {\rm Z} ({\rm X}) basis
as the sifted key, which we denote by ${\vec s}_{\rm A, \rm Z}$ (${\vec s}_{\rm A, \rm X}$) and ${\vec s}_{\rm B, \rm Z}$ (${\vec s}_{\rm B, \rm X}$), respectively. 
\item (Parameter estimation) Alice and Bob announce ${\vec s}_{\rm A, \rm X}$ and ${\vec s}_{\rm B, \rm X}$ over an authenticated public channel,
and they calculate the Hamming weight ${\sf wt}({\vec s}_{\rm A, \rm X}\oplus {\vec s}_{\rm B, \rm X})$, which is the number of mismatches between
${\vec s}_{\rm A, \rm X}$ and ${\vec s}_{\rm B, \rm X}$. From ${\sf wt}({\vec s}_{\rm A, \rm X}\oplus {\vec s}_{\rm B, \rm X})$, Alice and Bob
estimate the number of bits that need to be removed from ${\vec s}_{\rm A, \rm Z}$ and ${\vec s}_{\rm B, \rm Z}$ in the privacy amplification step.
\item (Error correction and privacy amplification) Alice and Bob perform error correction on ${\vec s}_{\rm A, \rm Z}$ and ${\vec s}_{\rm B, \rm Z}$
by using an error correcting code chosen from the family selected in Step (1). It requires the exchange of $\lambda_{\rm EC}$ bits of syndrome information over the authenticated public channel. Then, they perform privacy
amplification on the corrected sifted keys by using a hash function chosen at random
from the pre-agreed set of hash functions.
As a result, Alice (Bob) obtains the bit string ${\vec f}_{\rm A, Z}$ (${\vec f}_{\rm B, Z}$).
\item (Error verification) Alice and Bob select at random a hash function from the pre-agreed
set of hash functions established in Step (1), and they perform error
verification on ${\vec f}_{\rm A, Z}$ and ${\vec f}_{\rm B, Z}$. This process consumes at most
$\log_{2}2/\epsilon_{\rm c}$ bits of a previously shared key \cite{Wegman}. If the error verification fails, they discard all the data. Otherwise,
Alice's (Bob's) key is ${\vec f}_{\rm A, Z}$ (${\vec f}_{\rm B, Z}$).
\end{enumerate}
We have three remarks to make. First, $N_{\rm det}$ includes both the basis matched and mismatched events.
Second, in practical QKD protocols, to achieve a high repetition rate, Alice may want to send
single-photons before she obtains Bob's announcement
in Loop phase 2 regarding with the previous single-photons.
As a result, Alice continues to send pulses even if the number of the detection events
exceeds $N_{\rm det}^{\rm (Ter)}$, and Bob may obtain a number of detection events greater than than $N_{\rm det}^{\rm (Ter)}$.
In this case, we define the protocol such that Bob simply discards the events that exceed $N_{\rm det}^{th}$
(they are actually treated as the non-detected events), and Alice and Bob proceed as prescribed in the Actual protocol.
Finally, we also note that in the parameter estimation step Alice does not need to announce ${\vec s}_{\rm A, \rm X}$. This is so because
it suffices if one of the parties calculates ${\sf wt}({\vec s}_{\rm A, \rm X}\oplus {\vec s}_{\rm B, \rm X})$. However, 
we prefer to keep the Actual protocol as described above to show that the
disclosure of ${\vec s}_{\rm A, \rm X}$ does not compromise its security.

To analyze the security of the Actual protocol, we proceed as in \cite{Azuma2, Mizutani, LT, SP, Koashi, Hayashi} and we
consider a virtual protocol that is equivalent to the Actual protocol from Eve's
viewpoint. The virtual protocol is introduced in the next section.

\section{VIRTUAL PROTOCOL AND BRIEF REVIEW OF THE INSECURITY OF THE BB84 PROTOCOL BASED
ON RANDOM SAMPLING}\label{Virtual protocol}

In this section, we present a virtual protocol that is mathematically equivalent to the
Actual protocol described above and, therefore, we can use it for the security proof.
To prove its security, one can either use the security proof technique based on the
uncertainty principle \cite{Koashi} or the one based on the entropic uncertainty relation \cite{Renner}.
Most importantly, the virtual protocol provides a clear description of the state to
which one can apply the security proof. Indeed, the main purpose of this section is to
describe such a state. Moreover, in the second part of this section, we discuss the essential points that render the
BB84 scheme with iterative sifting insecure when the parameter estimation is based on the random sampling theory
\cite{iterative}.

We start by noting that Alice's state preparation process in the Actual protocol can
be decomposed into two steps. First, she prepares two single-photons in the
maximally entangled state $\ket{\phi^+}_{\rm AB}:=(\ket{0_{\rm Z}}_{\rm A}\ket{0_{\rm Z}}_{\rm B}+\ket{1_{\rm Z}}_{\rm A}\ket{1_{\rm Z}}_{\rm B})/\sqrt{2}$. Afterwards, she
measures the system ${\rm A}$ in a basis $\alpha\in\{{\rm Z}, {\rm X}\}$ selected probabilistically, and sends Bob the system
${\rm B}$. In the virtual protocol shown below, we assume that Alice adopts this
two-step process in her state preparation.

One important point in the virtual protocol is that we require that Alice and Bob
postpone their measurements for the sifted key generation, while we keep all the
classical and quantum information that is available to Eve exactly the same as that in
the Actual protocol. As for Alice, we simply consider that she postpones her
measurements on her systems ${\rm A}$ until the end of the Loop phases
(see virtual protocol below).
This is possible because whether she performs
her measurement before or after sending the system ${\rm B}$ does not change any
statistics. In the case of Bob, we exploit the basis independent detection efficiency
condition. This allows us to decompose his measurement into the following two
steps. First, Bob performs a filtering operation described by a pair of Kraus operators
$\{\sqrt{M_{\rm fail}}, \sqrt{\openone- M_{\rm fail}}\}$, where the former Kraus operator corresponds to the non-detected instance and the
latter one represents the detected instance. If the filtering operation outputs a detected instance,
then Bob performs the two-valued measurement on the resulting system (which is not necessarily a qubit) by using the
${\rm Z}$ or ${\rm X}$ basis, which he selects probabilistically. This measurement certainly supplies
Bob with the bit value for the chosen basis. In so doing, we consider the virtual
protocol running as follows.
\newline

(Virtual protocol)
\begin{enumerate}
\renewcommand{\labelenumi}{(\arabic{enumi})}
\item The same as Step (1) in the Actual protocol.
\item (Loop phase 1) Alice prepares two single-photons in the maximally entangled state $\ket{\phi^+}_{\rm AB}$, and sends Bob
the system ${\rm B}$ via the quantum channel.
\item (Loop phase 2) Bob performs on the incoming system ${\rm B'}$ the filtering operation $\{\sqrt{M_{\rm fail}}, \sqrt{\openone- M_{\rm fail}}\}$
to determine whether or not the pulse is detected, and he announces his result
over an authenticated public channel. If the pulse is detected, Alice and Bob increase $N_{\rm det}$ by 1 unit.
\item (Loop phase 3) Alice and Bob
probabilistically and independently select the basis, and they announce their choices over an authenticated public channel.
If both of them have selected the ${\rm Z}$ (${\rm X}$) basis, they keep systems A and B and define
them as the ${\rm Z}$ (${\rm X}$)- system. Moreover, if $N_{\rm det}=N_{\rm det}^{\rm (Ter)}$ is satisfied, Alice and Bob
proceed to Step (5), otherwise Alice and Bob return to Step (2).
\item (Sifted key generation) Alice and Bob measure all the {\rm Z} ({\rm X})-systems with the {\rm Z} ({\rm X}) basis. As a result,
they generate the sifted keys ${\vec s}_{\rm A, \rm Z}$ (${\vec s}_{\rm A, \rm X}$) and ${\vec s}_{\rm B, \rm Z}$ (${\vec s}_{\rm B, \rm X}$), respectively.
\item (Error correction and privacy amplification) The same as Step (6) in the Actual protocol.
\item (Error verification) The same as Step (7) in the Actual protocol.
\end{enumerate}

It is easy to see that this virtual protocol provides Eve exactly the same quantum and
classical information as the Actual protocol. Moreover, the statistics of Alice and
Bob's measurement outcomes do not change either. This means that we can consider
such a virtual protocol for the security proof. To prove the security of the virtual protocol, we need to find
a quantum state from which Alice and Bob generate the key. For that,
let $\rho^{\rm AB'}_{N_{\rm Z},{\vec I}}$ be the quantum
state of Alice and Bob's $N_{\rm Z}$ pairs of the {\rm Z}-systems after the termination of the Loop
phases. Note that $N_{\rm Z}$ depends on the classical information announced by Alice and Bob during the Loop phases.
This is so because Eve's attack might dependent of the previous basis announcements, but, importantly 
once the Loop phases finish then $N_{\rm Z}$ is a fixed number. Moreover, it also depends
on all the protocol parameters that Alice and Bob agreed in Step (1) such as the secrecy
parameter $\epsilon_{\rm s}$, and the quantity $ N_{\rm det}^{\rm (Ter)}$.
All such announcements and protocol parameters are represented by ${\vec I}$.

We are interested in the bit strings ${\vec s}_{\rm A, \rm Z}$ and
${\vec s}_{\rm B, \rm Z}$, which are the outcomes of the ${\rm Z}$ basis 
measurements on the systems in the state $\rho^{\rm AB'}_{N_{\rm Z},{\vec I}}$.
Importantly, as already mentioned above, the fact that we have a description of such a state before
the ${\rm Z}$ basis measurements to obtain the sifted keys enables us to employ 
both the security proof technique based on the uncertainty principle \cite{Koashi} and the one based on the entropic
uncertainty relation \cite{Renner}. 
If we employ the former (latter) one, the proof is applied to the state $\rho^{\rm AB'}_{N_{\rm Z},{\vec I}}$
($\rho^{\rm AB'E}_{N_{\rm Z}, {\vec I}}$ that is an extension of $\rho^{\rm AB'}_{N_{\rm Z},{\vec I}}$ that accommodates Eve's system).
In both cases, the security argument requires to consider Alice and Bob's fictitious ${\rm X}$ basis measurements
(instead of the ${\rm Z}$ basis measurements) on such a state. This fictitious ${\rm X}$ basis measurement is used to
define the phase error rate $e_{\rm ph}$, i.e., 
the error rate that Alice and Bob would have observed if they
had measured the systems in the state $\rho^{\rm AB'}_{N_{\rm Z},{\vec I}}$ using the ${\rm X}$ basis.
Note that if we choose to employ the security proof based on the uncertainty relation, it is conventional to use the fact that
the max-entropy is upper-bounded by the binary entropy of the phase error rate \cite{Renner}.
As we will explain in the next section, however, to estimate $e_{\rm ph}$ we will not directly use the state $\rho^{\rm AB'}_{N_{\rm Z},{\vec I}}$,
but we will consider a density operator $\rho^{\rm AB'}_{i|{\vec I}_{i-1}, {\vec x}_{i-1}}$ that represents a conditional state of the $i^{th}$ 
pair of the ${\rm Z}$-systems. The exact definition of this state is given in the next section.

If the phase error rate $e_{\rm ph}$ is estimated to be 
$e_{\rm ph} \le {\overline e_{\rm ph}}$ except with probability $\eta$, then it is known, for instance, that the security proof
based on the uncertainty principle \cite{Koashi, Mizutani, Hayashi} delivers a ($\epsilon_{\rm s}+\epsilon_{\rm c}$)-secure key of
length $l$ given by
\begin{eqnarray}\label{Key rate}
l\ge N_{\rm Z}\left[1-h({\overline e_{\rm ph}})\right]-\log_{2}\frac{2}{\epsilon_{\rm s}^2-\eta}-\lambda_{\rm EC}-\log_{2}\frac{2}{\epsilon_{\rm c}}\,.\ 
\end{eqnarray}
Here, $\epsilon_{\rm s}$ is the secrecy parameter, $\lambda_{\rm EC}$
is the number of bits exchanged during the error correction step, and $\epsilon_{\rm c}$ is the correctness parameter. A security
proof based on the entropic uncertainty relation provides almost the same key rate formula \cite{Renner}.

Since the phase error rate is not directly observed in the Actual protocol, one
typically estimates it from quantities like ${\vec s}_{\rm A, \rm X}$ and ${\vec s}_{\rm B, \rm X}$.
For this, a number of security proofs \cite{BB84 random} employ the theory of random sampling without replacement \cite{Serfling}.
In so doing, one usually regards the bit string ${\vec s}_{\rm A, \rm X}\oplus {\vec s}_{\rm B, \rm X}$
as the test bits to estimate the number of fictitious phase errors.
This is so because the positions of the bits ${\vec s}_{\rm A, \rm X}$ and ${\vec s}_{\rm B, \rm X}$ within the whole set of
detected instances seem to be chosen probabilistically and independently.

Unfortunately, however, it has been shown recently \cite{iterative} that the random sampling theory 
does not provide the correct estimation for $e_{\rm ph}$ when the protocol uses the basis dependent iterative
sifting. To understand why this is so, let us recall the necessary conditions to apply the random sampling theory (see, for instance, Lemma 7
in \cite{random sampling}). For this, we first review the mathematical
description of the theory, and then its application to the BB84 protocol with iterative sifting.
Suppose that we have a sequence ${\vec Y}:=(Y_{1}, Y_{2}, \cdots, Y_{n})$ of two-valued random variables where $Y_{a}$ ($a=1, 2, \cdots, n$)
takes a bit value. Then, suppose that we want to estimate the number of say $1$'s in a
code sequence that is randomly chosen from the $n$ bits, given that we know the
number of $1$'s in the rest of the sequence (which is called the test sequence). Let ${\vec T}:=(T_{1}, T_{2}, \cdots, T_{n})$
be the sequence of random variables that determines which bits form the test sequence.
That is, if $T_{a}=1$ ($T_{a}=0$) then the position $a$ is chosen as part of the test (code) sequence. Importantly, the random sampling theory
requires that (i) ${\vec T}$ has to be independent of ${\vec Y}$ and (ii) it has to be uniformly distributed, i.e., all the code sequences
with the same length are chosen equally likely.

It turns out, however, that none of these two conditions (i) and (ii) is satisfied in the case of the
BB84 with the basis dependent iterative sifting. Note that in this context ${\vec Y}$ corresponds to the sequence
of random variables that represent the phase error pattern, and ${\vec T}$ 
is the pattern of the basis choices. That is, $T_{a}$ determines whether the $a^{th}$ bit belongs to the code
sequence, i.e., Alice and Bob agree in the ${\rm Z}$ basis, or to the test sequence, i.e., Alice
and Bob agree in the ${\rm X}$ basis. Then, due to the iterative sifting procedure, it can be
shown that ${\vec T}$ is not independent of ${\vec Y}$. To prove this, note that Eve's decision of
whether or not she induces a phase error at the position $a$ can be made adaptively
based on all previous announcements done by Alice and Bob. That is, since $Y_{a}$ can
depend on the basis choice $(T_{1}, T_{2}, \cdots, T_{a-1})$, ${\vec T}$ is not independent of ${\vec Y}$.
What is worse is that, as was shown in \cite{iterative}, ${\vec T}$ is not uniform. That is, even if the choice of each of $T_{a}$ is made uniform,
there appears some bias in the probability distribution of ${\vec T}$ due to the termination condition. 
This means that the random sampling theory does not
guarantee the security of the BB84 with iterative sifting when the termination condition is basis dependent.

When the basis independent sifting procedure is employed, we have that, at least, condition (i) is not satisfied according to the same reasoning. 
As will be shown in the next section, to solve this problem, we use Azuma's inequality \cite{Azuma} and 
the fact that Eve cannot know Alice and Bob's basis choice beforehand.

\section{Phase error rate estimation based on Azuma's inequality}\label{Security proof}

In this section we show that the BB84 protocol with iterative sifting is secure if the
phase error rate is estimated using Azuma's inequality \cite{Azuma} and the termination condition is basis independent (but the disclosure of 
the basis information at each round is allowed).
As was already introduced above, recall that the phase error rate is the fictitious bit error
rate that Alice and Bob would observe if they measured the systems in the
state $\rho^{\rm AB'}_{N_{\rm Z},{\vec I}}$ using the ${\rm X}$ basis. In order to estimate this quantity, it is convenient to consider
a protocol where Alice and Bob indeed perform the ${\rm X}$ basis measurements on
$\rho^{\rm AB'}_{N_{\rm Z},{\vec I}}$. The protocol is a modified version of the Virtual protocol introduced above, but it terminates
when its Loop phases finish. More specifically, we replace the Loop phase 3 in the Virtual
protocol with the following step.

\begin{enumerate}
\renewcommand{\labelenumi}{(\arabic{enumi})}
\setcounter{enumi}{3} 
\item (Loop phase 3')
Alice and Bob probabilistically and independently select the basis, and they announce their choices over an authenticated public channel. 
Then, they measure systems ${\rm A}$ and ${\rm B'}$
using the ${\rm X}$ basis, and record their respective outcomes. If the termination condition $N_{\rm det}= N_{\rm det}^{\rm (Ter)}$ 
is satisfied, they obtain the $N_{\rm Z}$ ($N_X$) bit strings ${\vec s}_{\rm A, \rm Z}^{\rm Vir}$
(${\vec s}_{\rm A, \rm X}$) and ${\vec s}_{\rm B, \rm Z}^{\rm Vir}$ (${\vec s}_{\rm B, \rm X}$), respectively,
which correspond to the instances with the ${\rm Z}$ (${\rm X}$) basis agreement. Otherwise Alice and Bob return to Step (2) in the
Virtual protocol.
\end{enumerate}

On the one hand, we have that from Eve's viewpoint the Loop phases 3 and 3' are
indistinguishable because Alice and Bob's announcements coincide.
This means that $N_{\rm Z}$ and $N_{\rm X}$ in Loop phase 3' are equal to those in the
original Virtual and Actual protocols. On the other hand, Loop phases 3 and 3' differ
in that the latter requires Alice and Bob to measure the systems ${\rm A}$ and ${\rm B'}$
immediately after choosing their bases. Moreover, the performed
measurement is always in the ${\rm X}$ basis independently of their basis choice. Hence,
${\vec s}_{\rm A, \rm Z}^{\rm Vir}$ and ${\vec s}_{\rm B, \rm Z}^{\rm Vir}$ 
correspond to the bit strings that Alice and Bob obtain if they
announce the ${\rm Z}$ basis but perform the ${\rm X}$ basis measurement.
Importantly, the bit error rate between ${\vec s}_{\rm A, \rm Z}^{\rm Vir}$ and ${\vec s}_{\rm B, \rm Z}^{\rm Vir}$ coincides with the
phase error rate that Alice and Bob would have obtained if they had measured their respective systems in the state
$\rho^{\rm AB'}_{N_{\rm Z},{\vec I}}$ using the ${\rm X}$ basis in the original Virtual protocol.
This is so because once Eve has sent the system ${\rm B'}$ to Bob at each round of the Loop phases, it is too late for her to change
its state. That is, it does not matter when Alice and Bob
actually measure their systems, and we can assume in Loop phase 3' that they
measure them immediately after the confirmation of the detection event at each
round. 


For later convenience, let $\rho^{\rm AB'}_{ i|{\vec I}_{i-1}, {\vec x}_{i-1}}$ be Alice and Bob's state of the $i^{th}$ detected systems
before the ${\rm X}$ basis measurements in the Loop phase 3'. 
Note that when both Alice and Bob
choose the ${\rm Z}$ (${\rm X}$) basis, such a system was named the ${\rm Z}$ (${\rm X}$)-basis system in Step (4) of the Virtual protocol.
The state $\rho^{\rm AB'}_{ i|{\vec I}_{i-1}, {\vec x}_{i-1}}$ has
the following two crucial properties.
First, it is dependent on all the information
announced by Alice and Bob up to the $(i-1)^{th}$ round as well as on all the predetermined protocol parameters.
This includes, for instance, all the previous basis information and detection pattern, the
secrecy parameter $\epsilon_{\rm s}$, and the quantity $N_{\rm det}^{\rm (Ter)}$
for the termination condition.
All such information disclosed up to the $(i-1)^{th}$ round is represented by the subscript ${\vec I}_{i-1}$.
Importantly, this means that $\rho^{\rm AB'}_{ i|{\vec I}_{i-1}, {\vec x}_{i-1}}$
accommodates the effects of iterative sifting and the termination
condition, especially it takes into account attacks that exploit the basis dependency.
Moreover, this state can be correlated with all the previous outcomes
obtained by Alice and Bob's ${\rm X}$ basis measurements up to the $(i-1)^{th}$ round of the Loop phases, and those outcomes are 
represented by ${\vec x}_{i-1}$. 
The second property is that $\rho^{\rm AB'}_{ i|{\vec I}_{i-1}, {\vec x}_{i-1}}$ does not depend on Alice and Bob's basis choices on the
 $i^{th}$ round. This is because Alice and Bob announce their basis choices for the $i^{th}$ round
after Eve has already interacted with the $i^{th}$ pulse and sent it to Bob.

 Next, we estimate the number of phase errors that Alice and Bob obtain by measuring the systems in the state $\rho^{\rm AB'}_{ i|{\vec I}_{i-1}, {\vec x}_{i-1}}$ for
$i=1, 2, \cdots, N_{\rm det}^{\rm (Ter)}$. Before we explain the estimation method in detail, let us begin by presenting a sketch of the method.
Suppose that Alice and Bob repeat the Loop phase
steps many times until the termination condition is eventually satisfied.
Importantly, at this termination point, all the measurements have finished and the number of detected events
is fixed and equal to $N_{\rm det}^{\rm (Ter)}$, and it is independent of all the basis choices.
This is important for defining sequences of bounded Martingale random variables with a fixed length. 
For the estimation, we consider a particular round in the Loop phase, say the $i^{th}$ round, and
we derive a relationship, for the $i^{th}$ round, between the probability of having a phase error and the one of having 
an error for the X-basis measurement outcomes when both Alice and Bob choose this basis. 
Finally, given
such a relationship, we can employ Azuma's inequality to transform these
probabilities to the actual number of events. This way, we can estimate the phase
error rate from the number of errors ${\sf wt}({\vec s}_{\rm A, \rm X}\oplus {\vec s}_{\rm B, \rm X})$ 
between the strings ${\vec s}_{\rm A, \rm X}$ and ${\vec s}_{\rm B, \rm X}$. Importantly, Azuma's inequality already takes
into account any correlation possibly generated by Eve among the rounds.
This includes the most general correlations
that could arise if Alice sends
the single-photon pulses all in once and Eve applies a joint operation over all the single-photon pulses. Even in this general case,
the $i^{th}$ state can be represented by $\rho^{\rm AB'}_{ i|{\vec I}_{i-1}, {\vec x}_{i-1}}$. This is so because the way how the announcements
are made is still in a sequential manner, and Alice and Bob's measurements for the previous $(i-1)$ pairs of the systems remain the same.

Let us now explain the estimation method in detail. We note that the following discussion does
not assume any specific ${\vec x}_{i}$ and ${\vec I}_{i}$, and hence it holds for any ${\vec x}_{i}$ and ${\vec I}_{i}$.
Following the prescription described above, we need to consider the following two probabilities. 
The first one is the conditional probability that Alice and Bob choose the ${\rm Z}$ basis
but measure the systems in the conditional state $\rho^{\rm AB'}_{ i|{\vec I}_{i-1}, {\vec x}_{i-1}}$ along the ${\rm X}$ basis
and obtain the phase error. The second one is the conditional probability
that Alice and Bob choose and announce the ${\rm X}$ basis and use this basis to measure the systems in the conditional state
$\rho^{\rm AB'}_{ i|{\vec I}_{i-1}, {\vec x}_{i-1}}$ and they obtain the error in the ${\rm X}$ basis. 
Let us denote the former (latter) probability by
$P^{(i)}({\rm Ph}|{\vec x}_{i-1}, {\vec I}_{i-1})$ ($P^{(i)}({\rm X\,\,error}|{\vec x}_{i-1}, {\vec I}_{i-1})$).  
These probabilities are given by
\begin{eqnarray}\label{Prob}
P^{(i)}({\rm Ph}|{\vec x}_{i-1}, {\vec I}_{i-1})&=&q_{\rm Z}{\rm Tr}\left(\rho^{\rm AB'}_{ i|{\vec I}_{i-1}, {\vec x}_{i-1}} \Pi_{\rm err}^{\rm (X)} \right)
\nonumber\\
P^{(i)}({\rm X\,\,error}|{\vec x}_{i-1}, {\vec I}_{i-1})&=&q_{\rm X}{\rm Tr}\left(\rho^{\rm AB'}_{ i|{\vec I}_{i-1}, {\vec x}_{i-1}} \Pi_{\rm err}^{\rm (X)} \right)\,,\nonumber\\
\end{eqnarray}
where $q_{\rm Z}:=p_{\rm Z}^{\rm (A)}p_{\rm Z}^{\rm (B)}$, $q_{\rm X}:=p_{\rm X}^{\rm (A)}p_{\rm X}^{\rm (B)}$, and 
$\Pi_{\rm err}^{\rm (X)}:=\ket{0_{\rm X}}_{\rm A}\bra{0_{\rm X}}\otimes M_{{\rm X}, 1}+\ket{1_{\rm X}}_{\rm A}\bra{1_{\rm X}}\otimes M_{{\rm X}, 0}$ is the POVM element associated to an error in the ${\rm X}$ basis.
Note that we can use $q_{\rm Z}$ and $q_{\rm X}$ in these equations because $N_{\rm det}^{\rm (Ter)}$ is independent of Alice and Bob's basis choices, that is, the probabilities $q_{\rm Z}$ and $q_{\rm X}$ are independent of ${\vec x}_{i-1}$ and ${\vec I}_{i-1}$.
From Eq.~(\ref{Prob}), we can readily obtain a relationship for the $i^{th}$ pair between the probabilities as 
\begin{eqnarray}\label{Prob relation}
\frac{P^{(i)}({\rm Ph}|{\vec x}_{i-1}, {\vec I}_{i-1})}{q_{\rm Z}}=
\frac{P^{(i)}({\rm X\,\,error}|{\vec x}_{i-1}, {\vec I}_{i-1})}{q_{\rm X}}\,.\nonumber\\
\end{eqnarray}

Now that we have the relationship in terms of the probabilities, the next step is to convert this relationship to the one in terms of the numbers.
For this, first we take the summation over $i$ of the probabilities in Eq.~(\ref{Prob relation}) to obtain
\begin{eqnarray}
&&\frac{\sum_{i=1}^{N_{\rm det}^{\rm (Ter)}}P^{(i)}({\rm Ph}|{\vec x}_{i-1}, {\vec I}_{i-1})}{q_{\rm Z}}\nonumber\\
&=&\frac{\sum_{i=1}^{N_{\rm det}^{\rm (Ter)}}P^{(i)}({\rm X\,\,error}|{\vec x}_{i-1}, {\vec I}_{i-1})}{q_{\rm X}}\,.
\label{relationship}
\end{eqnarray}
Next, for any ${\vec I}_{i-1}$ and ${\vec x}_{i-1}$, we define the following random variables
\begin{eqnarray}
X_{{\rm Ph}, j}&:=&\Lambda_{{\rm Ph}, j}-\sum_{i=1}^{j}P^{(i)}({\rm Ph}|{\vec x}_{i-1}, {\vec I}_{i-1})\\\label{R.V for Ph}
X_{{\rm X\,\,error}, j}&:=&\Lambda_{{\rm X\,\,error}, j}-\sum_{i=1}^{j}P^{(i)}({\rm X\,\,error}|{\vec x}_{i-1}, {\vec I}_{i-1})\,,\nonumber\\
\label{R.V for Xerr}
\end{eqnarray}
where $\Lambda_{{\rm Ph}, j}$ ($\Lambda_{{\rm X\,\,error}, j}$) represents the number of instances among the first
$j$ detected events where both Alice and Bob choose the ${\rm Z}$ (${\rm X}$) basis,
measure their systems along the ${\rm X}$ basis, and obtain a phase error (bit error in the ${\rm X}$ basis).
For later use, we define $X_{{\rm Ph}, 0}=0$ and $X_{{\rm X\,\,error}, 0}=0$. 
Note that $j$ runs from $0$ to $N_{\rm det}^{\rm (Ter)}$.
Importantly, these random variables satisfy the bounded differential condition (BDC), i.e, $|X_{{\rm Ph}, j}-X_{{\rm Ph}, j-1}| \le1$
and $|X_{{\rm X\,\,error}, j}-X_{{\rm X\,\,error}, j-1}| \le1$ for any $1\le j\le N_{\rm det}^{\rm (Ter)}$. Also,
they are Martingale with respect to ${\vec x}_{j}$ and ${\vec I}_{j}$. 
Here, Martingale means that $\{X_{{\rm Ph}, j}\}_{j=0}^{N_{\rm det}^{\rm (Ter)}}$ and
 $\{X_{{\rm X\,\,error}, j}\}_{j=0}^{N_{\rm det}^{\rm (Ter)}}$ satisfy
\begin{eqnarray}
E[X_{{\rm Ph}, j+1}|{\vec x}_{j}, {\vec I}_{j}]&=&X_{{\rm Ph}, j}\,,\\
E[X_{{\rm X\,\,error}, j+1}|{\vec x}_{j}, {\vec I}_{j}]&=&X_{{\rm X\,\,error}, j}
\end{eqnarray}
for $j=0, \cdots, (N_{\rm det}^{\rm (Ter)}-1)$ and for any ${\vec x}_{j}$ and ${\vec I}_{j}$. That is, the expectation value of $X_{{\rm Ph}, j+1}$ conditional on ${\vec x}_{j}$ and ${\vec I}_{j}$ is $X_{{\rm Ph}, j}$, and similarly for $X_{{\rm X\,\,error}, j+1}$. We note that with BDC and
Martingale, we can apply Azuma's inequality to $\{X_{{\rm Ph}, j}\}_{j=0}^{N_{\rm det}^{\rm (Ter)}}$ and
 $\{X_{{\rm X\,\,error}, j}\}_{j=0}^{N_{\rm det}^{\rm (Ter)}}$ \cite{Azuma-detail}. This is what we do next. 

An important point to be able to apply Azuma's inequality to the sequence of the bounded Martingale random variables, $\{X_{{\rm Ph}, j}\}_{j=0}^{N_{\rm det}^{\rm (Ter)}}$,
is that the length of this sequence is fixed and independent of all the basis choices, which is clearly satisfied in our case thanks to the
basis independent termination condition. Therefore, we can directly apply Azuma's inequality \cite{Azuma-detail} to $\{X_{{\rm Ph}, j}\}_{j=0}^{N_{\rm det}^{\rm (Ter)}}$ to obtain that
$\forall \delta>0$, $\forall N_{\rm det}^{\rm (Ter)}>0$, $\forall{\vec I}_{i-1}$, $\forall{\vec x}_{i-1}$, and for all outcomes of $\Lambda_{{\rm Ph}, N_{\rm det}^{\rm (Ter)}}$ with
$N_{\rm det}^{\rm (Ter)}\ge\Lambda_{{\rm Ph}, N_{\rm det}^{\rm (Ter)}}\ge0$,
\begin{eqnarray}
&& {\rm Pr}\left(\Lambda_{{\rm Ph}, N_{\rm det}^{\rm (Ter)}}-\sum_{i=1}^{N_{\rm det}^{\rm (Ter)}}P^{(i)}({\rm Ph}|{\vec x}_{i-1}, {\vec I}_{i-1})\ge
N_{\rm det}^{\rm (Ter)}\delta\right)\nonumber\\
&\le& e^{-(N_{\rm det}^{\rm (Ter)})\delta^2/2}
\label{Azu1}
\end{eqnarray} 
holds \cite{Azuma, Azuma2, Mizutani, Azuma-detail}. Similarly, we have that $\forall \delta>0$, $\forall N_{\rm det}^{\rm (Ter)}>0$,
$\forall{\vec I}_{i-1}$, $\forall{\vec x}_{i-1}$, and for all outcomes
of $\Lambda_{{\rm X\,\,error}, N_{\rm det}}^{\rm (Ter)}$ with $N_{\rm det}^{\rm (Ter)}\ge\Lambda_{{\rm X\,\,error}, N_{\rm det}^{\rm (Ter)}}\ge0$
\begin{eqnarray}
{\rm Pr}\Bigg(&\sum_{i=1}^{N_{\rm det}^{\rm (Ter)}}&P^{(i)}({\rm X\,\,error}|{\vec x}_{i-1}, {\vec I}_{i-1})-\Lambda_{{\rm X\,\,error}, N_{\rm det}^{\rm (Ter)}}\nonumber\\
&\ge& N_{\rm det}^{\rm (Ter)}\delta\Bigg)\le e^{-(N_{\rm det}^{\rm (Ter)})\delta^2/2}\,.
\label{Azu2}
\end{eqnarray} 

Then, by combining Eqs.~(\ref{relationship}), (\ref{Azu1}), and (\ref{Azu2}) with the fact that $\Lambda_{{\rm X\,\,error}, N_{\rm det}^{\rm (Ter)}}={\sf wt}({\vec s}_{\rm A, \rm X}\oplus {\vec s}_{\rm B, \rm X})$, we have that
\begin{eqnarray}
\Lambda_{{\rm Ph}, N_{\rm det}^{\rm (Ter)}}&\le& \frac{q_{\rm Z}}{q_{\rm X}}{\sf wt}({\vec s}_{\rm A, \rm X}\oplus {\vec s}_{\rm B, \rm X})+\left(\frac{q_{\rm Z}}{q_{\rm X}}+1\right)N_{\rm det}^{\rm (Ter)}\delta
\nonumber\\
&=:&{\overline \Lambda_{{\rm Ph}, N_{\rm det}^{\rm (Ter)}}}
\end{eqnarray}
holds except for error probability less than $2e^{-(N_{\rm det}^{\rm (Ter)})\delta^2/2}$.
Finally, we set ${\overline e_{\rm ph}}={\overline \Lambda_{{\rm Ph}, N_{\rm det}^{\rm (Ter)}}}/N_{\rm Z}$ and
$\eta=2e^{-(N_{\rm det}^{\rm (Ter)})\delta^2/2}$, and
by plugging them into Eq.~(\ref{Key rate}), we have the final key length with the security parameter equal to $\epsilon_{\rm s}+\epsilon_{\rm c}$.
This concludes the security proof. 

\section{Security of the loss-tolerant protocol and Discussion}\label{Discussion}
We have demonstrated that the BB84 protocol with iterative sifting and the basis independent termination condition is secure if one
employs Azuma's inequality to estimate the parameters needed for
privacy amplification. The essential points of the security proof can be summarized
as follows. (1) First, we consider a virtual protocol where the measurements that
generate the sifted keys are postponed. This way, we obtain a clear description of the
state of the shared systems. (2) Second, this state satisfies that, at any particular
round within the Loop phase steps, it can only depend on the announcements made
before that round as well as on all the protocol parameters. (3) Third, we have that
this state is independent of Alice and Bob's basis choice at the current round, as the
basis announcement for each round is always made once Bob has received his
system. (4) Finally, the length of the random variable sequences of our interest is fixed
to a predetermined parameter and it is independent of all the basis choices.

Thanks to (1), we can use the conventional framework of security proofs such as, for
instance, the technique based on the uncertainty principle \cite{Koashi} or the one based on
the entropic uncertainty relation \cite{Renner}. Also, as we have seen above, point (2) does
not allow us to use the random sampling theory because this theory requires that the
state of the whole Alice and Bob's systems is independent of all their basis choices. And,
finally, thanks to (3) and (4) we can use Azuma's inequality for the estimation instead.
This is so because the state is independent of the basis choice at the current round and Alice and Bob
choose the bases with probabilities as prescribed in the protocol.
Note as well that Azuma's inequality already takes into account any possible
correlation among the events.

We note that several QKD protocols with iterative sifting and the basis independent
termination condition satisfy the conditions (2)-(4). Therefore if one uses Azuma's inequality \cite{Azuma2, Mizutani}
and a virtual protocol that satisfies the condition (1) in the security proof, such protocols are secure even with iterative sifting and the basis independent
termination condition. To see this, let us consider the loss-tolerant protocol \cite{LT}. 
This protocol differs from the BB84 protocol only in that the loss-tolerant protocol uses just three states (let us denote them as
$\ket{\phi_{1}}_{\rm B}$, $\ket{\phi_{2}}_{\rm B}$, and $\ket{\phi_{3}}_{\rm B}$) and it exploits the basis mismatch events.
Except for these differences, all the quantum and classical communication phases of the loss-tolerant protocol are the same as those in the 
BB84 protocol. In particular, the three states are associated to two bases like in the BB84 protocol, and therefore,
all the classical information announced during the Loop phases in the loss-tolerant protocol with iterative sifting and 
the basis independent termination condition is exactly the same as that of the BB84 protocol with the same sifting and
termination condition.

Now, let us see if the loss-tolerant protocol with iterative sifting and the basis independent
termination condition is secure by first checking whether (1)-(4) are satisfied or not.
As for (1), fortunately, the security proof presented in \cite{LT} considers an entangled state
essentially of the form $\sum_{i=1}^{5}\sqrt{P(i)}\ket{i}_{\rm A}\ket{\phi_{i}}_{\rm B}$, where $\{\ket{i}_{\rm A}\}$ is a set of orthonormal states
of a system to be held by Alice,  $P(i)$ is the probability
of sending the state $\ket{\phi_{i}}$, and $\ket{\phi_{4}}$ and $\ket{\phi_{5}}_{\rm B}$ are linear combinations of $\ket{\phi_{1}}_{\rm B}$
and $\ket{\phi_{2}}_{\rm B}$. Moreover, the key is generated from delayed measurements on the system ${\rm A}$ and the system ${\rm B'}$
where ${\rm B'}$ represents the system ${\rm B}$ after Eve's intervention. Therefore, (1) is satisfied.
Moreover, (2)-(4) are satisfied thanks to the aforementioned fact that all the classical information announced is
the same as the one of the BB84 at any particular round within the Loop phase. Therefore, all of (1)-(4) are satisfied.
Finaly, note that since the security proof in \cite{LT} employs Azuma's inequality to deal with any possible
correlation among the events, it is also valid for the correlation arising from the basis announcements from all the previous rounds \cite{specific}.
As a result, the loss-tolerant protocol is secure even if it is implemented with iterative sifting and the basis independent termination
condition.
This also means, for example, that a generalized version of the loss tolerant protocol \cite{in preparation} is also secure.


We remark that our proof employs the basis independent termination condition where
the termination of the Loop phases is based on the number of detection events regardless of the basis choices. 
We leave for our future work 
the security proof of a protocol with the basis dependent termination condition.

\section{Acknowledgment}
K. T. acknowledges support from CREST, JST, and H.-K. L. acknowledges financial support from NSERC, US ONR, CFI and ORF.
M.C. gratefully acknowledges support
from the Galician Regional Government (program ``Ayudas para proyectos de
investigacion desarrollados por investigadores emergentes,'' Grant No. EM2014/033, and consolidation of Research Units: AtlantTIC),
the Spanish Ministry of Economy and Competitiveness (MINECO), and the Fondo Europeo de
Desarrollo Regional (FEDER) through Grant No. TEC2014-54898-R, and the European Commission (Project QCALL).
A.M. acknowledges support from Grant-in-Aid for JSPS Fellows (KAKENHI Grant No. JP17J04177).

\bibliographystyle{apsrev}

\begin{thebibliography}{99}
\bibitem{footnote2}Here, we consider that Eve's interaction with the flying qubits is independent of the basis information at the
$i^{th}$ round; but it can be dependent on all the previous rounds. However, Eve's choice of measurement (to extract information from her quantum memory) can depend on 
the basis information from the $i^{th}$ round (as well as on any additional information obtained during the key distillation process).
\bibitem{footnote} A key advantage of Azuma's inequality
over random sampling theory is that Azuma's inequality applies to any bounded Martingale whereas the random sampling 
theory does not. Indeed, even if an eavesdropper adapts her eavesdropping strategy for 
future rounds of signals according to the publicly announced basis information for all previous 
rounds of signals from Alice and Bob, the random variables under consideration
still form a bounded Martingale. The fact that the state sent by Alice in the 
$i^{th}$ round is basis-independent (for the basis information in the $i^{th}$ round) is sufficient to guarantee 
that it is a "fair game" that Alice and Bob are playing with the eavesdropper, Eve, and that there is 
no way for Eve to enhance her winning strategy based on classical information (e.g. 
basis information) that she has acquired from previous rounds.
\bibitem{specific} More specifically, the operator ${\hat D}_{s_{X}}$ in \cite{LT}, which represents Eve's operation and Bob's
measurement, becomes dependent on all the previous announcements with the basis announcement, but it is still independent of the
basis choice at the current round. Mathematically, this dependency is the only one modification needed to prove the security of the
loss-tolerant protocol with iterative sifting and the basis independent termination condition. Noting that the security proof in \cite{LT} is 
valid for any ${\hat D}_{s_{X}}$ as long as this operator is independent of the basis choice at the current round, the loss-tolerant protocol with
iterative sifting and the basis independent termination condition is also secure.
\end{thebibliography}

\end{document}